\documentclass[11pt]{article}
\usepackage{latexsym}
\usepackage{graphicx}
\usepackage{amsfonts,amssymb}

\begin{document}

\title{The clustering coefficient and community structure of bipartite networks}
\author{Peng Zhang$^{1}$, Jinliang Wang$^{1}$, Xiaojia Li$^{1}$, Zengru Di$^{1}$, Ying Fan$^{1}$\footnote{Author for correspondence: yfan@bnu.edu.cn}, \\
\\ 1. Department of Systems Science, School of Management,\\
Center for Complexity Research,\\
Beijing Normal University, Beijing 100875, P.R.China }

\maketitle

\begin{abstract}
Many real-world networks display a natural bipartite structure. It
is necessary and important to study the bipartite networks by using
the bipartite structure of the data. Here we propose a modification
of the clustering coefficient given by the fraction of cycles with
size four in bipartite networks. Then we compare the two definitions
in a special graph, and the results show that the modification one
is better to character the network. Next we define a edge-clustering
coefficient of bipartite networks to detect the community structure
in original bipartite networks.
\end{abstract}



\section{Introduction}\label{introduction}

In recent years, as more and more systems in many different fields
can be depicted as complex networks, the research in complex
networks has been gradually becoming an important issue in the study
of complexity\cite{Review1,SIAM,Albert}. A network is composed of a
set of vertices and edges which represent the relationship between
two nodes. Examples include WWW, internet, food webs, biochemical
networks, social networks, and so
on\cite{PNAS,jazz,Williams,emaila,emailb,overlapping}. The research
in networks not only raises new concepts and methods, but also helps
us understand complex systems.

Many real-world networks display a natural bipartite structure, such
as the actors-films network\cite{movie}, the papers-scientists
network\cite{paper,paper1,paper2} and so on. In bipartite networks,
there are two kinds of nodes called top nodes and bottom nodes. The
edges only connect a pair of vertices belongs to different sets.
When we want to investigate some properties of them, we often
project them into one-mode networks which are also called classical
networks first. However, given the one-mode network of a bipartite
graph, it generally loses some information of the original bipartite
network, brings an inflation of the number of edges and other
drawbacks caused by projection\cite{drawback}. We believe that it
will affect the properties especially the community structure of the
networks. So we will pay more attention to study the community
structure and other properties of the original bipartite networks,
and develop some methods for detecting community structure in the
original bipartite networks.

Because of the drawbacks of projection, many authors try to analyze
the networks by using the bipartite structure of the data. Some
notions and properties, which are investigated in original bipartite
networks, are also introduced, such as clustering
\cite{stephen,original4}, overlap \cite{Phillip}, betweenness
\cite{stephen}, and others
\cite{stephen,drawback,Guler,original4,Garry}.

The outline of this article is as follows. In Section \ref{c4},
clustering coefficient, as one of the most important properties in
classical graphs, also attracts us much attention in bipartite
networks research. We propose a definition of clustering coefficient
based on the study in\cite{original4}. Then we use it to observe the
clustering coefficient of two real-world networks. In Section
\ref{community}, we use an algorithm based on the clustering
coefficient of links to detect the community structure of original
generated bipartite networks.In Section \ref{conclud} we give some
concluding remarks.

\section{The clustering coefficient of bipartite networks}
\label{c4}

The clustering coefficient $C_{3}$ is one of the most important
properties in classical networks. It defines the fraction of the
number of observed triangles to all possible triangles in networks.
It can be used to characterize the small-world networks\cite{movie},
understand the synchronization in scale-free networks\cite{McGraw},
and analyze networks of social relationships\cite{Newman,Holme}.
Refer to one-mode networks, the clustering coefficient of bipartite
networks should attract us much attentions\cite{original4}.

A bipartite network consists of two different kinds of nodes. The
links only can exist between two nodes which are from two distinct
sets. Many real-world networks display a natural bipartite
structure, such as the actors-films network, the papers-scientists
networks and so on. The clustering coefficient of classical graphs
measures the density of triangles. However, as the definition of
bipartite networks, the triangle can not be formed in it. The basic
clique in bipartite networks is a square. The clustering coefficient
$C_{4}$ should quantify the density of squares similar as the
clustering coefficient $C_{3}$. In social language, it calculates
the probability of that my friends have common friends except me.
Some other definitions of the clustering coefficient in bipartite
networks have been proposed\cite{stephen,Garry}. In this paper, we
present a new definition based on the one mentioned
in\cite{original4}.

In\cite{original4}, the clustering coefficient is defined as the
fraction of the number of observed squares to the total number of
possible squares in the graphs. For a given node $i$, the number of
observed squares is given by the number of common neighbors among
its neighbors, while the total number of possible squares is given
by the sum over each pair of neighbors of the product between their
degrees, after subtracting the common node $i$ and an additional one
if they are connected. The equation is:

\begin{equation}
C_{4,mn}(i)=\frac{q_{imn}}{(k_{m}-\eta_{imn})(k_{n}-\eta_{imn})+q_{imn}}.
\label{c4e}
\end{equation}
where $m$ and $n$ are a pair of neighbors of node $i$, and $q_{imn}$
is the number of squares which include these three nodes.
$\eta_{imn}=1+q_{imn}+\theta_{mn}$ with $\theta_{mn}=1$ if neighbors
$m$ and $n$ are connected with each other and $0$ otherwise.

\begin{figure}
\includegraphics[width=7cm]{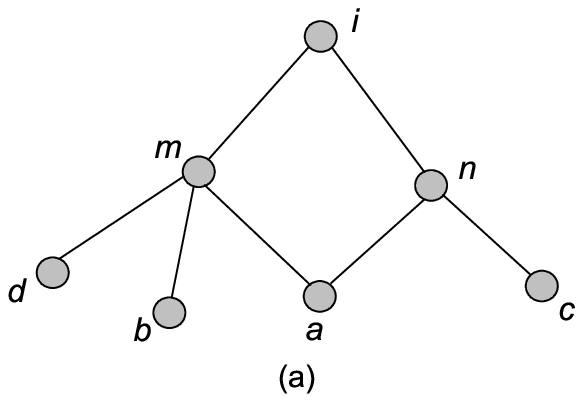}\includegraphics[width=7cm]{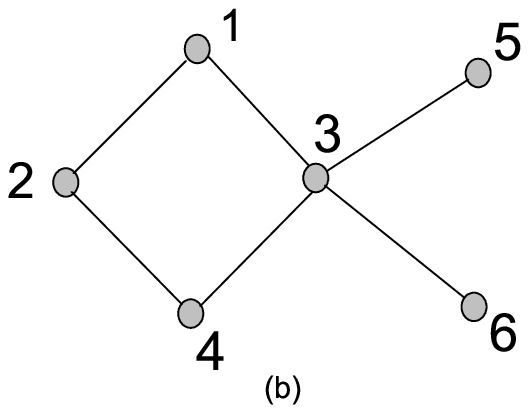}
\caption{(a) An example to show equation \ref{c4me} is better. (b) A
graph consists of six nodes}\label{fish}
\end{figure}

We thought that there is a drawback of considering the total number
of possible squares. The denominator of equation \ref{c4e} should be
changed into $(k_{m}-\eta_{imn})+(k_{n}-\eta_{imn})+q_{imn}$. The
equation is corrected as

\begin{equation}
C_{4,mn}(i)=\frac{q_{imn}}{(k_{m}-\eta_{imn})+(k_{n}-\eta_{imn})+q_{imn}}.
\label{c4me}
\end{equation}

where the representation of each parameter is the same as above.
Here we give an example to show that why we do this change. In
Figure \ref{fish} (a), considering the node $m$ and $n$, which are
the first neighborhoods of node $i$. It has $q_{imm}=1$ square and
$k_{m}=4$, $k_{n}=3$, $\theta_{mn}=0$. The denominator of equation
\ref{c4e} equals to $3$ in this case. It is not in accord with what
we see in figure. In figure \ref{test}, there should be $4$ possible
squares as $iman$, $imbn$, $imdn$ and $imcn$. Our definition of
equation \ref{c4me} gives a better answer. We also use these two
definitions to calculate the clustering coefficient of a special
graph (shown in Figure \ref{test} (b)). The results are shown in
table $1$. Taking the connections among vertices and the results
given by two equations into account, we consider that the
denominator definition of equation \ref{c4me} gives a better answer
to compute the clustering coefficient $C_{4}$. Because the distinct
connections of each node would cause different properties of each
node, especially the clustering coefficient.

\begin{center}
Table $1$ shows the clustering coefficient of each node in figure
\ref{fish} obtained by equation \ref{c4e} and \ref{c4me}.
\end{center}
\begin{center}
\begin{tabular}{|c|c|c|c|c|c|c|}

\hline

 & node 1 & node 2 & node 3 & node 4 & node 5 & node 6\\
\hline

eq 1 & 1 & 1 & 1 & 1 & 0 & 0\\
\hline

eq 2 & 0.3333 & 1 & 0.2 & 0.3333 & 0 & 0\\
\hline
\end{tabular}\label{compare12}
\end{center}

In order to complete the comparison between two definitions of the
clustering coefficient shown in equation \ref{c4e} and \ref{c4me},
we need to choose a lot of real bipartite-networks database. The
first one is the Econophysicists bipartite network built by
ourselves, which is composed of $818$ authors and $777$ papers. A
books-readers database obtained from Beijing Normal University
library during one semester, with $17593$ readers and $91750$ books.
The analysis results gotten by equation \ref{c4e} and \ref{c4me} are
displayed in table $2$ and $3$.

\begin{center}
Table $2$ displays the clustering coefficient of authors and papers
in the Econophysicists bipartite network obtained by equation
\ref{c4e} and \ref{c4me}.
\end{center}
\begin{center}
\begin{tabular}{|c|c|c|}

\hline

 & Authors & Papers\\
\hline

eq 1 & 0.18923 & 0.14353\\
\hline

eq 2 & 0.305 &0.16916\\
\hline
\end{tabular}\label{comparescientist}
\end{center}

\begin{center}
Table $3$ shows the clustering coefficient of books and readers in
the books-readers network obtained by equation \ref{c4e} and
\ref{c4me}.
\end{center}
\begin{center}
\begin{tabular}{|c|c|c|}

\hline

 & Books & Readers\\
\hline

eq 1 & 0.00063 & 0.00321\\
\hline

eq 2 & 0.00449 & 0.00632\\
\hline
\end{tabular}\label{comparebook}
\end{center}

\section{The community structure of bipartite graphs }
\label{community}

Different metrics of connections strength among vertices form the
community structure. Community structure is the groups of network
vertices. Within groups there are dense internal links among nodes,
but between groups nodes loosely connected to the rest of the
network\cite{Girvan}. It is one of the most important characters to
understand the functional properties of complex structures. Recent
empirical studies on networks display that there are communities in
most social and biology networks\cite{Girvan,metabolic}. This
finding is very significant to understand network structure. Taking
collaboration network of jazz musicians for an example, the analysis
reveals the presence of communities which have a strong correlation
with the recording location of the bands, and also shows the
presence of racial segregation between the musicians\cite{jazz}. In
food web, communities reveal the subsystem of
ecosystem\cite{Williams}. Email network can be divided into
departmental groups whose work is distinct and the communities
reveal organization structure or the results reveal the
self-organization of the network into a state where the distribution
of community sizes is self-similar\cite{emaila,emailb}. The deep
research in community structure will make us comprehend and analyze
the characteristic of systems better.

All above works are done in one-mode networks. However, many
real-world networks display a natural bipartite structure, such as
the actors-films network, the papers-scientists networks and so on.
When we want to investigate the community structure of them, we
often project it into one-mode network first. We believe that the
projection will bring some drawbacks and affect the properties
especially the community structures of the networks. So we should
pay more attention to analyze it in original bipartite graphs.

\begin{figure}
\center \includegraphics[width=9cm]{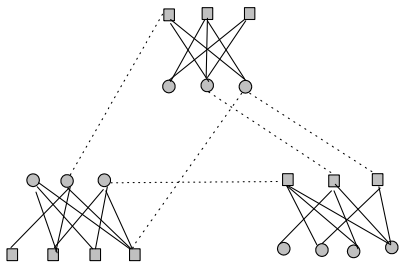} \caption{A
schematic representation of a bipartite network with community
structure. There are three communities of dense internal links
(solid lines), with sparse connections (dot lines) between them
}\label{example}
\end{figure}

Similar to classical networks, community structure of bipartite
networks is the groups of nodes. Within groups there are dense
internal links among two different sets of nodes, but between groups
nodes loosely connected to ones belonged to the other set of the
network (shown in Fig. \ref{example}). To the one-mode networks,
$Filippo$ $Radicchi$ $et al$ have proposed a divisive
algorithm\cite{edgecluster}. They considered the edge-clustering
coefficient, defined in analogy with the usual node-clustering
coefficient. Here we also can define the edge-clustering coefficient
of bipartite networks, as the number of squares to which a given
edge belongs,divided by the number of squares that might potentially
include it. For the edge-connecting top node $i$ to bottom node $j$,
the edge-clustering coefficient is

\begin{equation}
C_{ij}=\frac{\sum_{m=1}^{k_{i}}{\sum_{n=1}^{k_{j}}{q_{ijmn}}}}{\sum_{m=1}^{k_{i}}{\sum_{n=1}^{k_{j}}{\theta_{ijmn}}}+(\sum_{m=1}^{k_{i}}k_{m}-1)+(\sum_{n=1}^{k_{j}}k_{n}-1)-\sum_{m=1}^{k_{i}}{\sum_{n=1}^{k_{j}}{q_{ijmn}}}}.
\label{cij}
\end{equation}
where $m$ is a neighbor of node $i$, and $n$ is one of $j$'s
neighbors. $q_{ijmn}=1$ if neighbors $m$ and $n$ are connected with
each other and $0$ otherwise. $\theta_{ijmn}$ is opposite to
$q_{ijmn}$. $k_{m}$ is the degree of node $m$. This algorithm works
as the GN algorithm, however, the edge with the smallest value of
$C_{ij}$ should be cut at each step.

Similar to test the performance of a method in one-mode networks,
here we apply the edge-clustering coefficient algorithm to a
computer-generated bipartite networks. The generated network is made
up of 64 top nodes and 64 bottom nodes. All the nodes are divided
into four separate communities. There are $16$ top nodes and $16$
bottom nodes in every community. Vertices are assigned to groups and
are randomly connected to vertices of the same group by an average
of $<k_{intra}>$ links and to vertices of different groups by an
average of $<k_{inter}>$ links. The degree of all vertices is fixed,
namely $<k_{intra}>+<k_{inter}>=16$. It is obvious that with
$<k_{inter}>$ increasing, the communities become more diffuse and it
becomes more difficult to detect the communities.

\begin{figure}
\includegraphics[width=7cm]{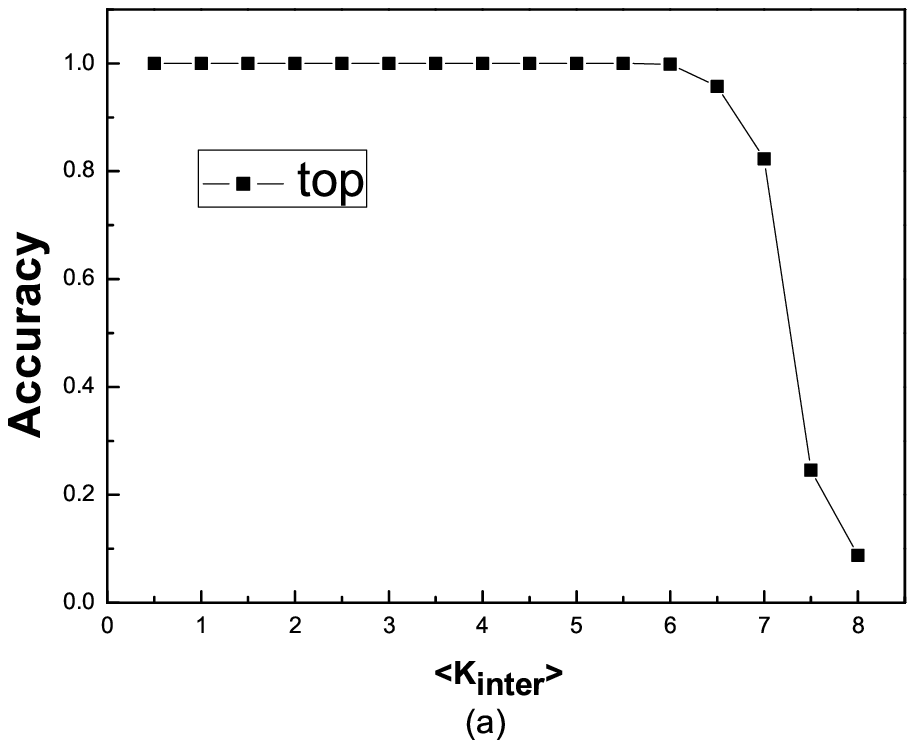}\includegraphics[width=7cm]{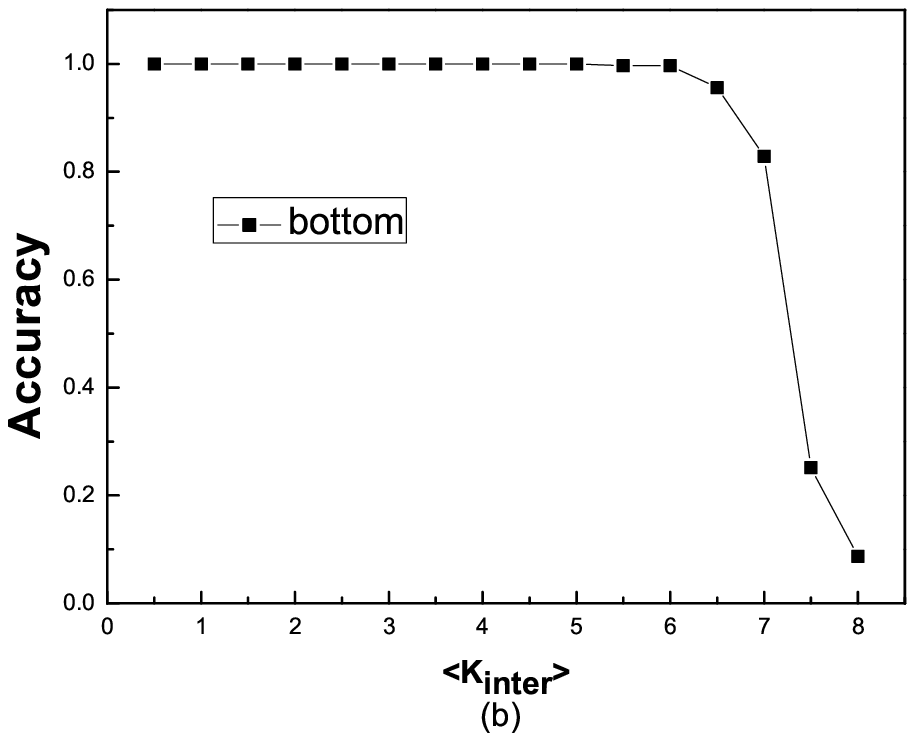}
\caption{Our algorithm performance as applied to computer-generated
bipartite networks with $n=128$ and four communities of $16$ top
nodes and $16$ bottom nodes each. Total average degree is fixed to
$16$. (a) is the accuracy of top nodes using computer-generated
bipartite networks with presumed community structure. (b) is the
accuracy of bottom nodes using computer-generated bipartite networks
with presumed community structure. The $x$-axis is the average of
connections between nodes in different groups $<k_{inter}>$.}
\label{book}
\end{figure}

In the following numerical investigations, we get $20$ realizations
of computer-generated bipartite networks under the same condition.
Based on these results, using the similarity function $S$ which has
been mentioned in\cite{zhangpeng}, comparing each divided groups
with presumed community structure. We get the accuracy of our
algorithm (shown in Fig. \ref{book} (a) and \ref{book} (b)).

\begin{figure}
\center \includegraphics[width=9cm]{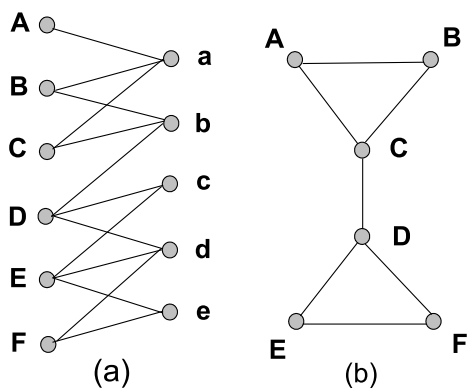} \caption{(a)
It is composed of $6$ top nodes and $5$ bottom nodes. (b)It is the
projection on top nodes of (a)}\label{bipartitegood}
\end{figure}

Here is a bipartite network, which includes $6$ top nodes and $5$
bottom nodes (shown in Fig. \ref{bipartitegood} (a)). According to
the definition of community structure of bipartite networks which we
mentioned in the beginning of this section, Fig. \ref{bipartitegood}
(a) is consist of three communities, as
\{\{A\},\{B,C,a,b\},\{D,E,F,d,e\}\}. As before, when we get a
bipartite network, first we often project it into a one-mode
network. Here we project this bipartite network into top nodes with
weights (shown in Fig. \ref{bipartitegood} (b)). It is divided into
two parts by using the WEO algorithm\cite{Duch}, as
\{\{A,B,C\},\{D,E,F\}\}. It is different from what is shown in the
graph. Next we use our algorithm to analyze the community structure
of this bipartite network, the result we get is same as what we see
from the graph.

\section{Conclusions}
\label{conclud}

In this paper, we proposed a modification of the clustering
coefficient given by the fraction of cycles with size four in
bipartite networks based on the work of $Pedro\ G.\ Lind\ et\ al$
\cite{original4}. We use these two definitions to calculate the
clustering coefficient of a special graph, and got that there is
difference between two results. We considered that the one we
defined gives a better answer with the distinct connections of each
node of graph. Then we discussed the community structure of
bipartite graphs, and defined an algorithm based on the
edge-clustering coefficient of bipartite networks. In this way, we
avoided the drawbacks and effects brought by the projection to the
analysis of community structure, just as the example we gave in the
end of section \ref{community}. At last, we tested the accuracy of
this algorithm in the computer-generated bipartite networks. We
found that when community structure is well defined by topological
linkage, it works well. But this algorithm only considered the nodes
which connected more than twice. This needs to be modified in the
future.

\section{Acknowledgement}
This work is partially supported by 985 Projet and NSFC under the
grant No.70771011, No.70431002 and No.60534080.

\end{document}